\documentclass[10pt,twocolumn,english,preprint,prl,amsfonts,amssymb,byrevtex,tightenlines,nobalancelastpage]{revtex4}
\usepackage[T1]{fontenc}
\usepackage[latin9]{inputenc}
\usepackage{graphicx}
\usepackage{amssymb}

\makeatletter

\newcommand{\lyxdot}{.}

\@ifundefined{textcolor}{}
{%
 \definecolor{BLACK}{gray}{0}
 \definecolor{WHITE}{gray}{1}
 \definecolor{RED}{rgb}{1,0,0}
 \definecolor{GREEN}{rgb}{0,1,0}
 \definecolor{BLUE}{rgb}{0,0,1}
 \definecolor{CYAN}{cmyk}{1,0,0,0}
 \definecolor{MAGENTA}{cmyk}{0,1,0,0}
 \definecolor{YELLOW}{cmyk}{0,0,1,0}
 }

\providecommand{\U}[1]{\protect\rule{.1in}{.1in}}

\makeatother

\usepackage{babel}

\begin{document}
Phase Diagram of Electrostatically Doped SrTiO$_{3}$

\author{Yeonbae Lee, Colin Clement, Jack Hellerstedt, Joseph Kinney, Laura
Kinnischtzke, S.D. Snyder, and A. M. Goldman}

\affiliation{School of Physics and Astronomy, University of Minnesota, 116 Church
St. SE, Minneapolis, MN 55455, USA}
\begin{abstract}
Electric double layer transistor configurations have been employed
to electrostatically dope single crystals of insulating SrTiO$_{3}$.
Here we report on the results of such doping over broad ranges of
temperature and carrier concentration employing an ionic liquid as
the gate dielectric. The surprising results are, with increasing carrier
concentration, an apparent carrier-density dependent conductor-insulator
transition, a regime of anomalous Hall effect, suggesting magnetic
ordering, and finally the appearance of superconductivity. The possible
appearance of magnetic order near the boundary between the insulating
and superconducting regimes is reminiscent of effects associated with
quantum critical behavior in some complex compounds. 
\end{abstract}
\maketitle
Strontium Titanate (STO) is a band insulator which has been rendered
conducting and superconducting at sufficiently low temperatures by
introducing oxygen defects, by doping with Nb, Zr, La, or Ta\cite{Koonce,Binnig,Hulm,Suzuki,Olaya,Gurvitch,Leighton},
through charge transfer processes resulting from the deposition of
several layers of LaAlO$_{3}$\cite{Reyen}, by electrostatic charging
using a conventional field effect transistor (FET) configuration \cite{Nakamura},
and by charging using an electric double layer transistor (EDLT) configuration
employing a polymer electrolyte\cite{Ueno}. In this letter we report
the results of an electrostatic charging experiment on SrTiO$_{3}$
using an ionic liquid (IL) in an EDLT configuration, which has revealed
unexpected features in the phase diagram as a function of temperature
and carrier concentration, resembling those in the phase diagrams
of unconventional superconductors, or systems exhibiting quantum critical
behavior \cite{Taillefer,SI}. 

EDLTs are configurations which perform in a manner similar to conventional
FETs. They contain a gate electrode, a semiconductor or insulator
into which charge can be accumulated or depleted, source and drain
electrodes, and for purposes of characterization, electrodes for measuring
longitudinal and transverse voltages relative to the current direction.
In this instance the gate dielectric is an IL, although polymer electrolytes
have also been employed. ILs are molten salts consisting of large
ions such that their Coulomb interaction is sufficiently small to
make them room temperature liquids. They have previously been used
to modify the conductivity of ZnO, and ZrNCl, revealing a metal-insulator
transition in ZnO, and superconductivity in ZrNCl\cite{Misra,Shimotani,Ye,Hongtao Yuan}.
Upon applying a gate voltage, ions move to the surface of the semiconductor
or insulator, forming an electric double layer (EDL) such that with
the charge induced in the channel, acts as a capacitor of nanoscale
thickness. A schematic of the operation of an EDLT is shown in Fig.
1. With ILs, charge transfers of order 10$^{14}$ cm$^{-2}$ or higher
are possible, opening up the possibility of significant modification
of the electronic properties of materials\cite{Ahn}. The advantage
of using ILs over almost most solid gate dielectrics is a greatly
enhanced charge transfer for the same gate voltage.

\begin{figure}
\includegraphics[scale=0.5]{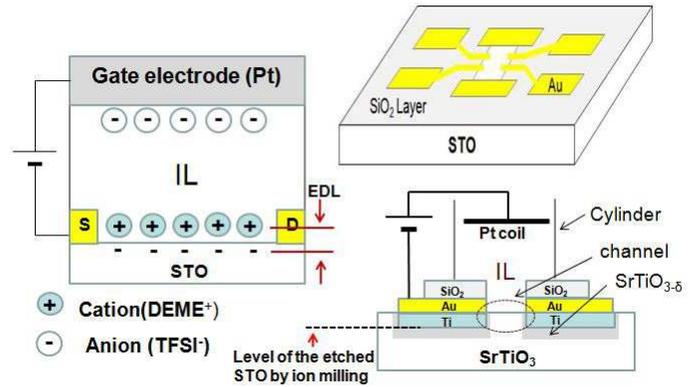} \caption{(color online) Schematic diagrams of the sample and the operation
of an EDLT: (Left) Upon application of a positive gate bias, anions
are attracted to the gate electrode and cations are repelled, this
creates an EDL at the sample surface. (Right) Schematics of top and
side views of the sample design. A Pyrex glass cylinder was glued
onto the STO wafer using a common epoxy to serve as an IL container.
Oxygen deficient SrTiO$_{3-\delta}$ was created below and around
the Ti/Au electrodes, but not across the channel.}

\end{figure}

Pristine STO (100) single-crystals were obtained from Princeton Scientific
Corp. Their surfaces were treated with buffered hydrofluoric acid
for 30 seconds and annealed at 780$^{\circ}$C for 6 hours to yield
an atomically flat surface terminated by TiO$_{2}$\cite{Kawasaki}.
Electrodes in a typical Hall bar geometry along with a four-probe
configuration for measuring longitudinal resistance were pre-patterned
by means of photolithography as shown in Fig. 1. Prior to the deposition
of Ti/Au electrodes (each layer, 70 nm in thickness), the STO was
ion milled to a depth of 70nm to create a layer of oxygen-deficient
metallic SrTiO$_{3-\delta}$. The depth of milling of the STO is shown
in Fig. 1. Milling was carried out to insure ohmic contact between
the Ti/Au electrodes and the surface layer of STO that would be electrostatically
gated into a conductive state\cite{Reagor}. Next a 100nm thick layer
of SiO$_{2}$ was deposited through a photo-patterned mask onto the
STO to define a conducting channel $140\mu m\times70\mu m$ in area.
The IL used here was N,N-diethyl-N-(2-methoxyethyl)-N-methylammonium
bis (trifluoromethyl sulphonyl)-imide (DEME-TFSI). This particular
IL has produced very high charge accumulations on the surface of semiconductors
such as ZnO with induced sheet carrier densities ($n_{s}$) of up
to $8\times10^{14}cm^{-2}$ being reported \cite{Hongtao Yuan}. Moreover
the ions of this IL are mobile down to 200K. The ability to apply
a gate voltage,$V_{g}$, at this relatively low temperature enables
us to raise voltage $V_{g}$ as high as 5V without significant and
irreversible chemical degradation that would occur at higher temperatures.

\begin{figure}
\includegraphics[scale=0.28]{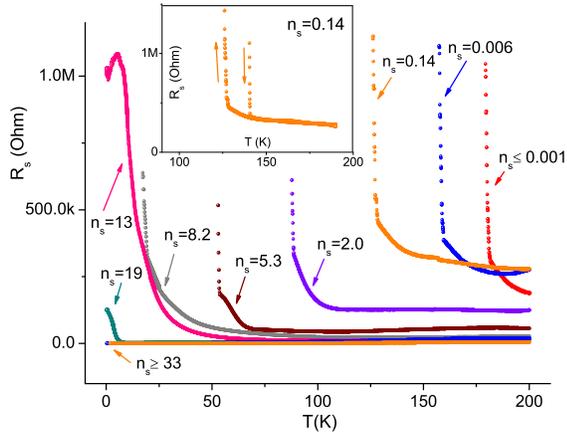}

\caption{(color online) Sheet resistance ($R_{s}$) vs. temperature ($T$)
on cooling, at various sheet carrier densities ($n_{s})$ (in units
of $10^{13}cm^{-2}$, obtained from the Hall effect at 180K). When
$n_{s}\geqq3.3\times10^{14}cm^{-2}$, the resistance is well below
the quantum resistance $h/e^{2}=25.8k\Omega$, and $dR_{s}$/$dT$>0
persists down to the base temperature of 0.35K. Note that the resistance
peak when $n_{s}=1.3\times10^{14}cm^{-2}$ shifted upon the application
of a magnetic field (see Fig. 5 ). An anomalous Hall effect was observed
at this and nearby charge transfers. The inset shows the hysteretic
behavior of the sharply rising curve with $n_{s}=1.4\times10^{12}cm^{-2}$.
Similar behavior was observed at other carrier concentrations. The
arrows in the inset indicate the temperature sweep direction.}

\end{figure}

Four-probe measurements were employed to determine the resistance
of samples. A Quantum Design Physical Property Measurement System
(PPMS) equipped with a $^{3}$He refrigerator insert enabled us to
vary the temperature between 300K and 0.35K and the magnetic field
between 0T and 9T. As the IL, DEME-TFSI, undergoes a rubber-like phase
transition below 240K and becomes a glass-like solid below 200K, where
the ionic mobility falls to zero\cite{Hongtao Yuan}. $V_{g}$ was
always changed at 230K. and ranged from 0V to 3V. It was applied using
a Keithley electrometer (Model 6517A), and was held constant throughout
subsequent cooling and measurement. 

The temperature dependencies of the sheet resistances $R_{s}$, of
an STO crystal at various values of $n_{s}$ are plotted in Fig. 2.
Undoped STO exhibited no measurable conductivity, and only with $V{}_{g}$
in excess of 1.8V was it possible to measure nonzero conductivity.
Within a given cycle of cooling and warming at a fixed value of $V_{g},$
there was no more than a 5 to 10\% shift in resistance when the crystal
was brought back to 230K. In a set of independent measurements using
this IL, the value of the capacitance in a parallel plate configuration
exhibited a small time dependence, owing to the complex dynamics of
the IL in its glassy state\cite{Rivera,Jacob Stevenson}. Areal carrier
density, $n_{s}$, rather than $V{}_{g}$, is the intrinsic parameter.
Values of $n{}_{s}$ were obtained from Hall effect data at 180K,
which was measured in magnetic fields $B$, of up to 5T, and $n_{s}$
ranged from $1\times10^{10}cm^{-2}$ to $8\times10^{14}cm^{-2}.$
The highest value of $n_{s}$ ever induced was $1.5\times10^{15}cm^{-2}$.
The areal carrier concentration, $n_{s}(V_{g})$, was not linear in
$V_{g}$, which is probably a consequence of the voltage dependence
of the IL's dielectric properties. 

In most plots of $R_{s}(T)$, we observed a sharp resistance rise
at a temperature which depended upon $n_{s}$ which was not instrumental\@.
This could be due to the reduction of carriers thermally excited from
states within the band gap, when decreasing the temperature. Such
an effect is believed to be responsible for the temperature dependence
of the voltage threshold for conductance as has been found in various
oxide FETs\cite{Taisuke}. However, the sharpness of the rise in $R_{s}(T)$,
and the fact that it is hysteretic, as shown in the inset of Fig.
2, strongly suggest that this behavior to be the signature of a first
order phase transition in the layer that is electrostatically doped. 

It is well known that chemically doped STO, exhibits superconductivity
at relatively low bulk carrier densities $n_{bulk}\sim10^{20}cm^{-3}$\cite{Koonce}.
Recently EDLT configurations were used to induce superconductivity
in an undoped insulating STO\cite{Ueno}. In the present work, we
were able to produce a very similar result albeit with a different
dielectric. The onset of a superconducting transition was observed
at a temperature of about 0.4K at $n_{s}=3.9\times10^{14}cm^{-2}$(measured
at 0.5K). Superconductivity either disappears or is just shifted to
lower temperatures, no more accessible with our set up, as we increased
the value of $n_{s}$. Data exhibiting the onset of superconductivity
is shown in Fig. 3.

\begin{figure}
\includegraphics[scale=0.6]{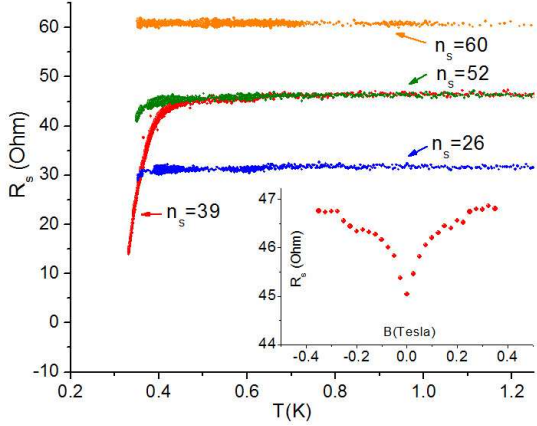}\caption{(color online) The onset of the superconductivity at various values
of $n_{s}$ (in units of $10^{13}cm^{-2},$ taken at 0.5K). Note that
at higher carrier densities the transition was not observed at temperatures
down to 350mK. In the inset, the $B$-dependence of $R_{s}$ at T
= 400mK for $n_{s}=3.9\times10^{14}cm^{-2}$ shows an increasing resistance
until saturation above 0.3T. This supports the observation of superconductivity.}

\end{figure}

\begin{figure}
\includegraphics[scale=0.3]{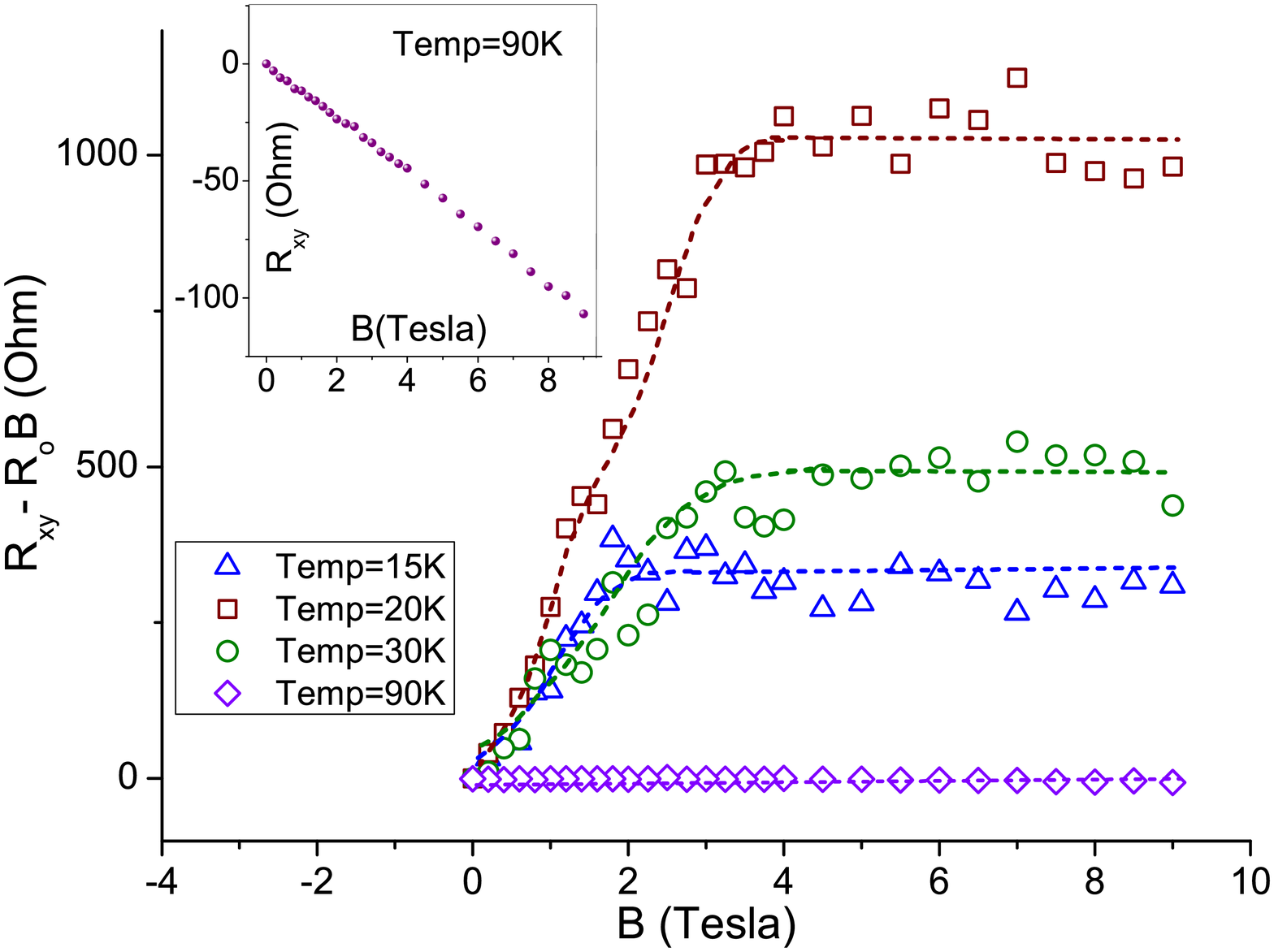}\caption{(color online) Anomalous Hall effect. This data was obtained with
$n_{s}=1.3\times10^{14}cm^{-2}$ (at 180K) over a temperature range
from 15K to 90K. The AHE effect was not observed at temperatures of
90 K and above, as shown in the inset (The inset shows the original
data before the OHE subtraction and it exhibits no signature of AHE.)
A similar effect was observed at neighboring values of $n_{s}$. }

\end{figure}

In Fig. 4. we plot $R_{xy}-R_{0}B$ vs. $B$, the Hall resistance
$R_{xy}$ vs. $B$ with the ordinary Hall Effect (OHE) term, $R_{0}B$
subtracted. This linear term was obtained from a fit to the high field
limit with $R_{0}=1/(en_{s})$. A deviation from the linearity of
the Hall resistance vs. magnetic field, known as anomalous Hall effect
(AHE), is observed. Similar effects have been reported at the SrTiO$_{3}$/LaAlO$_{3}$
interface \cite{BenShalom,Snir Seri,C. Bell}. The AHE can result
from a multiple band structure\cite{C. Bell} and/or a magnetization
induced by external field \cite{BenShalom,Snir Seri}. The observation
that the Hall resistance changes at low magnetic fields and saturates
at some high field, favors the magnetization scenario. The AHE effect
was not observed at temperatures of 90 K and above, as shown in the
inset of Fig. 4. %
\begin{figure}

\includegraphics[scale=0.28]{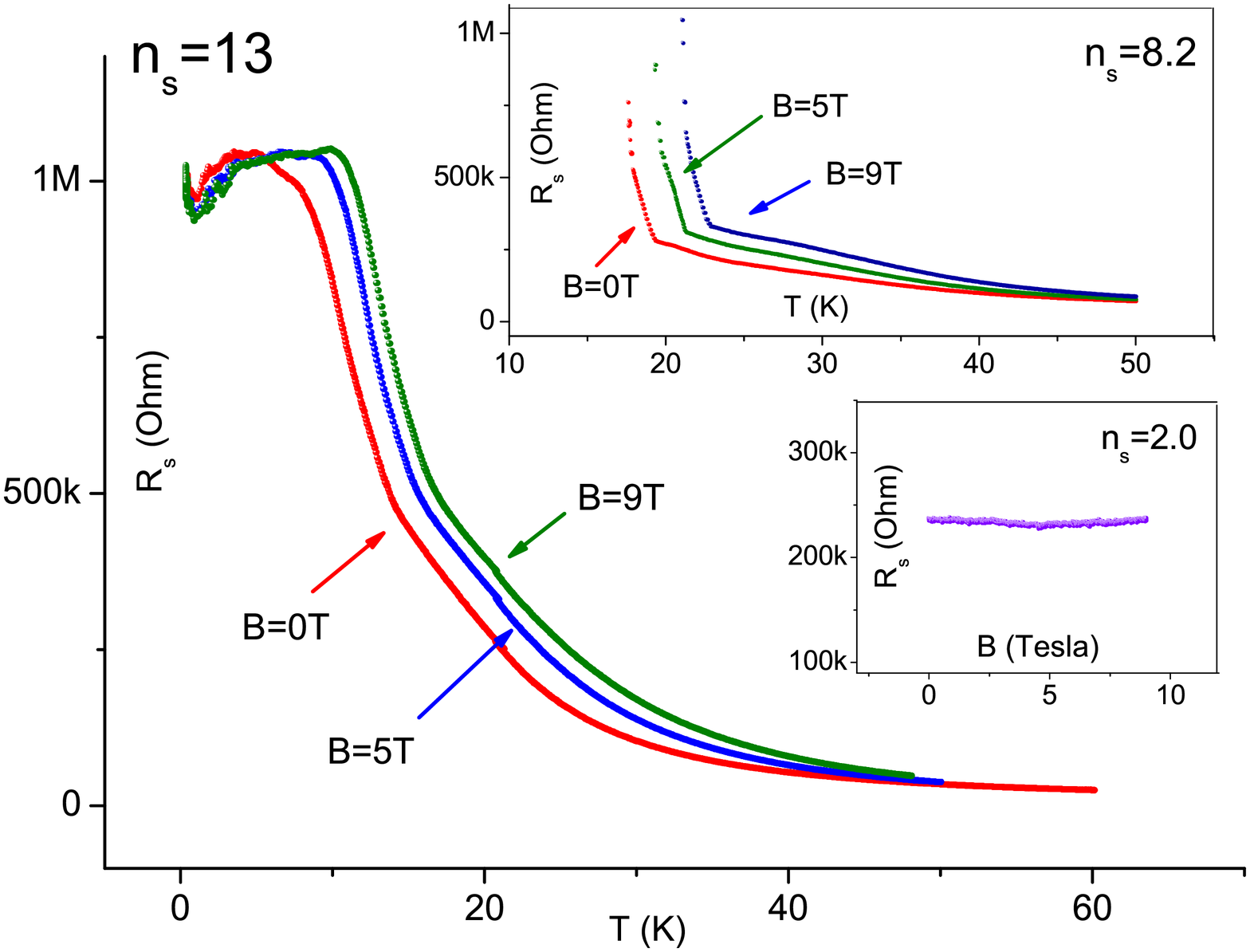}\caption{(color online) $R_{s}$ vs. T at various magnetic field up to 9 Tesla
with $n_{s}$ = 13 x 10$^{13}$ cm$^{-2}$with the latter obtained
at 180K. The lower-right inset shows $R_{s}\left(B\right)$ at $n_{s}=2.0\times10^{13}cm^{-2}$and
$T=$ 95K, which is the onset temperature for the sharp rise in $R_{s}\left(T\right)$
shown in Fig 2. It exhibits no change with field.}

\end{figure}

The magnetic field-dependence of $R_{s}$ in the region where an AHE
was observed was also investigated. Examples are shown in Fig. 5.
The peak at $n_{s}=1.3\times10^{14}cm^{-2}$ shifted to higher temperatures
upon the application of a magnetic field and the response to field
was not hysteretic. Similar behavior was found at the somewhat lower
carrier density of $n_{s}=8.2\times10^{13}cm^{-2}$ (the upper-right
inset of Fig. 5). This effect disappeared for carrier densities of
$2.0\times10^{13}cm^{-2}$ and lower (lower-right inset). We suspect
this effect is associated with magnetic order. It should be noted
that ferromagnetic order of the electron spins within the Ti\textit{-3d}
conduction band was predicted for both LaTiO$_{3}$/SrTiO$_{3}$ and
LaAlO$_{3}$/SrTiO$_{3}$ interfaces\cite{Pentcheva,Okamoto} and
evidence for this was presented by A. Brinkman \textit{et al}.\cite{Brinkman}
in the form of hysteretic magnetoresistance curves. In the present
work there is no hysteresis. Consequently it is not certain that ferromagnetism
is responsible for the observed phenomena. 

Lastly, we have drawn a tentative phase diagram in the space of $T$
and $n_{s}$, summarizing our observations (Fig. 6). Due to the limited
temperature range of the PPMS, it is not certain that there exists
a dome-like feature for the variation of the superconducting transition
temperature with $n{}_{s}$, analogous to the result found for chemically-doped,
bulk STO\cite{Koonce}. 

In summary, we have found a series of different behaviors of electrostatically
doped STO with the doping carried out using an EDLT configuration.
We have documented the systematic decrease of the temperature of carrier
freeze-out with increasing carrier concentration, with this apparent
conductor-insulator transition disappearing when superconductivity
turns on. The appearance of a peak in $R(T)$, which is magnetic field
dependent, in the vicinity of this crossover to superconductivity,
along with an anomalous Hall effect suggests the possibility of a
magnetic phase in the crossover regime. This is reminiscent of what
is found near quantum critical points in strongly correlated electron
systems\cite{SI}. Lower temperature measurements may elucidate the
precise behavior. An open question is the actual depth of penetration
of the induced charge layer. Recent theoretical work suggests that
it may be localized within a very narrow layer on the order of one
lattice constant rather than some greater distance\cite{Shklovskii}.
If that is indeed the case then the properties revealed in this work
need not be analogous to those found in bulk doping experiments.%
\begin{figure}
\includegraphics[scale=0.3]{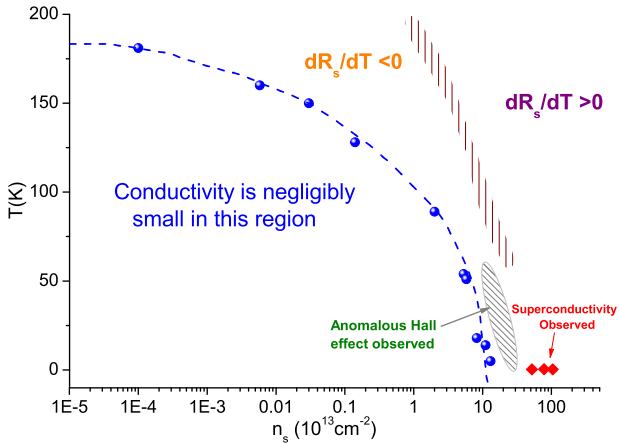}\caption{(color online) Phase diagram of STO: the plot of $T$ vs. $n_{s}$
(taken at 180K) shows the features found. The various dashed lines
are simply guides to the eye. The dots represent temperatures and
values of $n_{s}$ at which R$_{s}$ appeared to diverge. The vertically
hatched regions correspond the approximate boundary where $dR_{s}/dT$
changes sign, and the diamonds denote the onset of superconductivity,
which does not change much, over a significant range of values of
$n_{s}$. The other crosshatched region denotes the range of temperatures
and carrier concentrations over which an AHE was found. }

\end{figure}

We would like to thank Yen-Hsiang Lin and Christopher Leighton for
fruitful discussions and advice. This work was supported by the NSF
under Grant No. NSF/DMR-0854752 and by the DOE under grant No. ED-FG02-02ER46004.
Part of this work was carried out at the University of Minnesota Characterization
Facility, a member of the NSF-funded Materials Research Facilities
Network via the MRSEC program, and the Nanofabrication Center which
receive partial support from the NSF through the NNIN program.

\end{document}